\documentclass[review,3p]{elsarticle}
\usepackage{graphics}
\usepackage{amsmath}
\usepackage{balance}
\usepackage{amssymb}
\usepackage{caption}
\usepackage{multirow}
\usepackage{booktabs}
\usepackage{color}
\usepackage[bottom]{footmisc}
\usepackage{epstopdf}
\usepackage{threeparttable}
\usepackage{bm}
\usepackage{tabularx}
\usepackage{booktabs}
\usepackage{balance}

\biboptions{square,sort&compress}

\journal{Acta Materialia}

\begin{document}

\title{Stacking fault energy of face-centered cubic metals: thermodynamic and \emph{ab initio} approaches}

\author[kth]{Ruihuan Li}
\author[turku,kth]{Song Lu\corref{song}}
\cortext[song]{Corresponding author: Song Lu, Tel: +46 8 7906215, lusommmg@hotmail.com}
\author[kth]{Dongyoo Kim}
\author[kth]{Stephan Sch\"onecker}
\author[dalian1,dalian2]{Jijun Zhao\corref{zhao}}
\cortext[zhao]{Corresponding author: Jijun Zhao, zhaojj@dlut.edu.cn}
\author[korea]{Se Kyun Kwon}
\author[kth,upp,hung]{Levente Vitos}
\address[kth]{Applied Materials Physics, Department of Materials Science and Engineering, Royal Institute of Technology, Stockholm SE-10044, Sweden}%
\address[turku]{Department of Physics and Astronomy, University of Turku, FI-20014 Turku, Finland}%
\address[dalian1]{School of Physics and Optoelectronic Technology and College of Advanced Science and Technology, Dalian University of Technology, Dalian 116024, China}%
\address[dalian2]{Key Laboratory of Materials Modification by Laser, Electron, and Ion Beams (Dalian University of Technology), Ministry of Education, Dalian 116024, People's Republic of China}%
\address[korea]{Graduate Institute of Ferrous Technology, Pohang University of Science and Technology, Pohang 37673, Korea}%
\address[upp]{Department of Physics and Astronomy, Division of Materials Theory, Uppsala University, Box 516, SE-75120, Uppsala, Sweden}%
\address[hung]{Research Institute for Solid State Physics and Optics, Wigner Research Center for Physics, Budapest H-1525, P.O. Box 49, Hungary}%

\date{11 November 2015}

\begin{abstract}
The formation energy of the interface between face-centered cubic (fcc) and hexagonal close packed (hcp) structures is a key parameter in determining the stacking fault energy (SFE) of fcc metals and alloys using thermodynamic calculations. Often the contribution of the planar fault energy to the SFE has the same order of magnitude as the bulk part, and thus the lack of a precise information about it can become the limiting factor in thermodynamic predictions. Here, we differentiate between the actual interfacial energy for the coherent fcc(111)/hcp(0001) interface and the ``pseudo-interfacial energy'' that enters the thermodynamic expression for the SFE. Using first-principles calculations, we determine the coherent and pseudo- interfacial energies for six elemental metals (Al, Ni, Cu, Ag, Pt, and Au) and for three paramagnetic Fe-Cr-Ni alloys. Our results show that the two interfacial energies significantly differ from each other. We observe a strong chemistry dependence of both interfacial energies. The calculated pseudo-interfacial energies for the Fe-Cr-Ni steels agree well with the available literature data.
\end{abstract}

\maketitle

\section{Introduction}

In metals with face-centered cubic structure, dislocations can dissociate into two Shockley partial dislocations connected by a faulted ribbon. The most often observed fault is the intrinsic stacking fault (ISF). Assuming infinite separation between the two partials, an ideal ISF is obtained by removing a single close packed fcc (111) layer from the perfect fcc matrix. The stacking fault energy $\gamma$ (SFE) is the excess energy needed to form an ISF. SFE has been recognized as an important parameter controlling the mechanical properties of fcc metals and alloys. The magnitude of the SFE determines the width of the partial dislocations and thus is of primary importance in many aspects of plasticity related to the dislocation mediated behaviors. The twining induced plasticity (TWIP) mechanism has been associated with the SFE. According to semi-empirical correlations, \cite{Cooman2011} small SFE favors twinning, whereas large SFE leads to narrowly dissociated or undissociated dislocations and thus dislocation glide is favored. Very small or negative SFE is known to be responsible for the transformation induced plasticity (TRIP) mechanism. The critical SFE values separating the TRIP/TWIP/slip regimes in engineering alloys vary significantly. \cite{Allain2004158,Saeed-Akbari20093076,Nakano2010167} One possible reason behind these uncertainties is the typical large errors associated with the available SFE data.

The stacking fault energy is an intrinsic material property that can in principle be measured by carefully designed experiments. However, being a very small energy (usually of order of 10-100 mJm$^{-2}$) the accurate determination is very difficult and the reported experimental data often carries very large error bars. \cite{Vitos2008,schramm1975,rhodes1977} During the last decades several theoretical methods have been developed and employed in establishing more reliable SFE databases. A very powerful method is based on thermodynamic approach. In the popular model proposed by Olson and Cohen \cite{Olson1976} (also named as volumetric model), the stacking fault is treated as a two-layers embryo with the hexagonal close packed structure embedded in the fcc matrix. Accordingly, the SFE of an infinitely large stacking fault may formally be expressed as

\begin{equation}\label{eq1}
\gamma = 2 \rho \Delta G^{\rm hcp-fcc}+2\sigma^{*},
\end{equation}
where $\Delta G^{\rm hcp-fcc}$ is the difference in Gibbs free energies of equilibrium hcp and fcc phases, $\rho$ is the molar surface density (mol m$^{-2}$) of the (111) atomic layer. $\sigma^{*}$ is usually considered as the fcc/hcp interfacial energy and the $2\sigma^{*}$ term in Eq. (\ref{eq1}) accounts for the two interfaces between the hcp embryo and the fcc matrix. This interfacial energy may however differ significantly from the coherent fcc/hcp interfacial energy ($\sigma$) since in the above model of the stacking fault the hcp embryo has only two layers and the two fcc/hcp interfaces are likely to interact with each other. Additionally, in the thermodynamic calculations, hcp phase is often assumed at equilibrium state, while in the coherent interface calculations, the hcp structure may have a quite different c/a due to the coherent strain. In that respect, in Eq. (\ref{eq1}) all differences between the true stacking fault and the ``embryo'' model are included in the interfacial energy $\sigma^{*}$.~\cite{Olson1976} Because of that in the following we refer to $\sigma^*$ as the pseudo-interfacial energy to distinguish it from the true coherent interfacial energy $\sigma$ of fcc/hcp.

No direct measurements of $\sigma^*$ are available. Instead, various efforts have been put forward to estimate $\sigma^*$ indirectly via Eq. (\ref{eq1}) utilizing the measured SFE. For homogeneous Fe-Cr-Ni alloys, Olson and Cohen~\cite{Olson1976} using the calculated Gibbs free energies and the measured stacking fault energies~\cite{Lecroisey1972} found $\sigma^{*}=10-15$ mJm$^{-2}$. Recently, Pierce \emph{et al.}~\cite{Pierce2014} measured the compositional dependence of the stacking fault energy of Fe-Mn based alloys (Fe-22/25/28Mn-3Al-3Si wt.\%) and calculated $\sigma^*$ according to Eq. (\ref{eq1}) using the thermodynamic Gibbs energies. They showed that $\sigma^*$ ranges form 8 to 12 mJ m$^{-2}$ in Fe-Mn-Al-Si alloys and from 15 to 33 mJ m$^{-2}$ in binary Fe-Mn alloys. It was emphasized that $\sigma^*$ exhibits a strong dependence on the difference in Gibbs energies of the fcc and hcp phases. In particular, the Gibbs energy of the hcp phase, which has no precise description due to its thermodynamical instability, significantly influences the estimated $\sigma^*$.~\cite{Nakano2010167} In practice, the pseudo-interfacial energy in thermodynamic calculations is often treated as an adjustable parameter to bring the theoretical SFE values in line with the observed deformation mechanisms.~\cite{Saeed-Akbari20093076} Because of these difficulties, the effectiveness of the thermodynamic SFE models is limited.

\begin{figure}[t]
	\begin{center}
		\includegraphics[width=8cm]{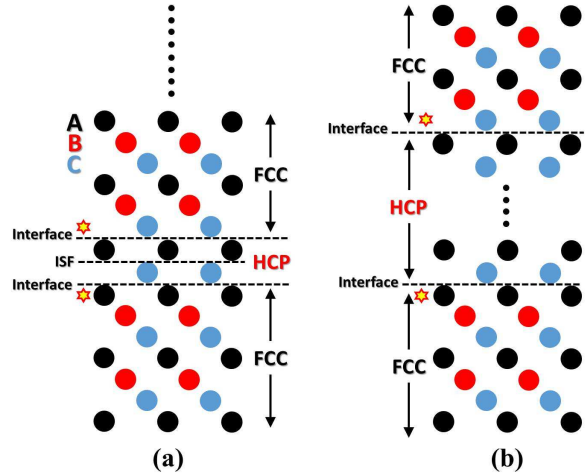}
		\caption{Stacking sequences and local structures for an hcp precipitate embedded in an fcc matrix. Panel (a): one stacking fault in fcc matrix which produces two hcp layers. Panel (b): $n$ stacking faults producing an hcp phase with 2$n$ layers. Asterisks mark the critical layers that might belonging to both hcp and fcc phases.}
	    \label{schematic1}
    \end{center}
\end{figure}

\emph{Ab initio} quantum mechanical modeling is an alternative approach to determine the SFE. Recently, it was employed to study the SFE in transition metal alloys and stainless steels.\cite{Lu20115728,Vitos2006,Vitos2006acta,lu2012} First-principles methods were also used to compute the interfacial energies. Following the volumetric model, one can increase the thickness of the hcp layers embedded in the fcc matrix from 2, corresponding to one ISF, to $2n$ describing $n$ consecutive stacking faults or equivalently two fcc/hcp interfaces separating the fcc and hcp parts of the layered system. A similar model was recently adopted by Rosalie \emph{et al.}, when studying the fcc/hcp phase interface in Al-Ag-Cu alloys.~\cite{Rosalie2014224} To the first order approximation, whether an atomic layer belongs to fcc or hcp phase depends on the stacking sequences of its two nearest neighboring layers. Namely, a layer having the two nearest neighbors with identical stacking sequence belongs to the hcp phase, otherwise it belongs to the fcc phase. This is illustrated schematically in Figure~\ref{schematic1}. In panel (a), showing the situation for $n=2$, we have marked the positions of the hcp/fcc interfaces when the system is considered as an hcp embryo embedded in fcc phase; and the position of stacking fault when the system is considered as fcc phase with one stacking fault. At the interfaces there are critical layers (marked by asterisk) which might erroneously be considered as belonging to the hcp phase. However, these layers have two different nearest neighbor layers (corresponding to local fcc packing) and thus should not be taken as being energetically equivalent to the hcp layers.~\cite{Rosalie2014224} Lee \emph{et al.} recently calculated the interfacial energy of fcc Fe(111)/hcp Fe(0001) using \emph{ab initio} methods.~\cite{Lee2012} They obtained an unreasonable interfacial energy (-241 mJ m$^{-2}$) compared to the commonly cited values (5-27 mJ m$^{-2}$). However, carefully examining their interfacial structure, one finds that they placed the hcp/fcc interface at ``...ABCABC$|$ABAB...'' instead of the correct position ``...ABCABCA$|$BAB...''. Therefore one fcc layer was erroneously taken as hcp layer, which resulted in the calculated ``interfacial energy'' containing an additional energy difference between the fcc and hcp phases scaled by the interface area $A$, i.e., $(E_{\rm fcc}-E_{\rm hcp})/A$.

The above analysis is solely based on the stacking sequence from the structural point of view. There is no composition difference between the two sides of the interface. In cases where the composition of the precipitate deviates from that of the matrix, it may look straightforward to define the phase interface as the composition interface. However, the situation could be more complicated, such as in the case of the hcp/fcc interface in Al-Ag alloys. Recent study showed that for precipitates with thickness of 1-3 stacking faults, the thickness of the Ag-enriched region is considerably wider than the size of the hcp region indicating that the phase interface is more likely a structural interface in the Ag-enriched alloys. For precipitates with $\geq 6$ stacking faults, the Ag-enriched zone was approximately equal to the width of the hcp region, indicating that the interface is a combination of structural and compositional interfaces.~\cite{Rosalie2014224} The interphase energies of hcp/fcc interfaces in Ag-Al alloys were studied by Finkenstadt \emph{et al.} to understand the large aspect ratio experimentally found for hcp precipitates.~\cite{Finkenstadt2010} The interfacial energy they got was 15-10 mJ m$^{-2}$. The interfaces were placed between pure fcc Al and hcp phase which is composed of alternating Al/Ag layers. However, the two interfaces in their model supercell were not symmetric and did not follow the definition of the structural interface between fcc and hcp phases, as discussed above.

In the present work, we study the coherent interfacial energies for the fcc(111)/hcp(0001) interfaces and the pseudo-interfacial energies in homogeneous metals and alloys. We employ an \emph{ab initio} total energy method that can account for alloying effects, so that the approach can easily be extended to solid solutions. Our approach avoids the ambiguous alloying configuration that might lead to segregation or partitioning. We calculate the interface energies for pure fcc metals (Al, Ni, Cu, Ag, Pt and Au) and a few Fe-Cr-Ni alloys which represent the main building block of austenitic stainless steels. The coherent interfacial energy is calculated with respect to the number of hcp layers and the pseudo-interfacial energy is calculated according to the thermodynamic expression for SFE (Eq. (\ref{eq1})) using the calculated SFE values by the supercell model. We demonstrate that there is a significant difference between the true and the pseudo-hcp/fcc interfacial energies. The large discrepancies originate from the actual definitions based on different reference structures. The rest of the paper is divided into two main sections and a conclusion. In Section \ref{method}, we introduce the models used for calculating the interfacial energies and give the calculation details. The results are presented and discussed in Section \ref{results}. The paper ends with conclusions.

\section{Methodology}\label{method}

In the present work, we employ two methods for calculating the stacking fault energy and the interfacial energy of the fcc(111)/hcp(0001) interface. The first method is based on a supercell model and provides a direct access to the interfacial energy as a function of layer thickness. The second approach maps the total energy into layer-layer interaction parameters offering a more straightforward insight into the mechanisms governing the interfacial energy in metals and alloys. This latter approach may be related to the thermodynamic model for calculating the SFE.

\subsection{Supercell model}\label{superm}

To calculate the coherent interfacial energy, we build a series of supercells containing fcc/hcp interfaces with various number of fcc (111) and hcp (0001) layers. The supercells are denoted by \textit{m}fcc+\textit{n}hcp or $(m,n)$, where \textit{m} and \textit{n} are the numbers of fcc and hcp layers respectively. Periodic boundary conditions are applied.
Considering the real situation that the lateral lattice constants are set up by fcc matrix, the stacking fault layers or hcp structure are distorted to form coherency with the fcc lattice. Hence, when there is a sizable difference between the equilibrium volumes of the fcc and hcp structures or the equilibrium hcp lattice has $c_{\rm hcp}/a_{\rm hcp}$ far away from the ideal value ($\sqrt{8/3}$), the embedded hcp layers in the supercells will be distorted (the interlayer distance being different from the equilibrium value ($\lambda_{\rm hcp}^ {0}$)). We therefore first relax the hcp structure with the constrain of $a_{\rm hcp}=a_{\rm fcc}^{111}=\sqrt{2}/2a_0$ ($a_0$ being the fcc lattice parameter) and find the interlayer distance $\lambda_{\rm hcp}$. Then the interlayer distances for ``bulk hcp'' inside the supercell ($\lambda_{\rm hcp}$), except the topmost ones ($\lambda_{\rm hcp}^{\rm topmost}$) which are the closest to the fcc/hcp interface, are fixed to $\lambda_{hcp}$. The interlayer distances between the fcc(111) layers are kept as the equilibrium values, $\lambda^0_{\rm fcc}=\sqrt{3}/3a_0$. The topmost hcp interlayer distance $\lambda_{\rm hcp}^{\rm topmost}$ and the interface separation between the fcc and hcp parts of the supercell $\lambda_{\rm fcc/hcp}$ were relaxed (see below).

Using the total energy $F_{\rm fcc/hcp}(m,n)$ of the \textit{m}fcc+\textit{n}hcp supercell, the interfacial energy may be calculated as

\begin{equation}\label{eq2}
		\sigma(m,n)=\frac{F_{\rm fcc/hcp}(m,n)-mF^0_{\rm fcc}-nF^{\rm SC}_{\rm hcp}}{2A},
\end{equation}
where $F^0_{\rm fcc}$ is the equilibrium total energy of the fcc lattice (per atom). $F^{\rm SC}_{\rm hcp}$ is the total energy of the constrained hcp lattice (per atom), which usually is larger than the actual equilibrium hcp energy $F^0_{\rm hcp}$. $A=\sqrt{3}/4a^2_0$ is the area of interface and the factor of 2 stands for the two interfaces present in the supercell. For a large enough $m$, the interfacial energy should converge with increasing $n$ towards a constant value representing the coherent fcc/hcp phase interface energy ($\sigma$).

For $n=0$, the interfacial energy is obviously zero. For $n=2$, the supercell has two hcp layers which can also be viewed as one intrinsic stacking fault. Hence, the intrinsic stacking fault energy may be computed from the present $(m,2)$ supercell energy as

\begin{equation}\label{eq3}
		\gamma(m)=\frac{F_{\rm fcc/hcp}(m,2)-(m+2)F^0_{\rm fcc}}{A}.
\end{equation}
With increasing fcc layer thickness $(m\rightarrow \infty)$ the above SFE converges to the true SFE $\gamma$.

In the present application, the number of fcc layers was chosen to be 9 ($m=9$) which was found to be large enough to remove the interaction due to the periodic boundary conditions. The number of hcp layers, on the other hand, was increased from 0 to 8 with an increment of 2. The interfacial energy is usually a very small quantity and it is very sensitive to the numerical details used in the \emph{ab initio} calculations. In order to keep the numerical noises at minimum, the fcc energy entering Eq. (\ref{eq2}) was calculated from the energy of 9fcc+0hcp supercell as $F^0_{\rm fcc}=F_{\rm fcc/hcp}(9,0)/9$. The energy of the distorted hcp phase was evaluated from the difference between the total energies of the $(9,8)$ and $(9,6)$ supercells, $F^{\rm SC}_{\rm hcp}=[F_{\rm fcc/hcp}(9,8)-F_{\rm fcc/hcp}(9,6)]/2$. We notice that computing $F^{\rm SC}_{\rm hcp}$ from the $(9,8)$ and $(9,4)$ supercells leads to very similar results as those presented here.

We studied the effect of relaxation of the interface separation between the fcc and hcp parts of the supercell $\lambda_{\rm fcc/hcp}$ and found that the fully relaxed layer separation is very close to the average value of the interlayer separations of fcc(111) and hcp(0001) layers. Therefore, in order to reduce the computational load, for all supercells we used $\lambda_{\rm fcc/hcp}\approx (\lambda^0_{\rm fcc}+\lambda_{\rm hcp})/2$. We also noticed that relaxing the topmost hcp interlayer distance $\lambda_{\rm hcp}^{\rm topmost}$ is important especially when $n$ is small.

\subsection{Axial Interaction Model}\label{aimm}

The stacking fault energy and the interfacial energy may be estimated using of the axial interaction model (AIM)~\cite{Vitos2006acta, Denteneer1987, Cheng1988}. Within this model, along the [111] direction, a particular stacking sequence of the close-packed (111) planes can be mapped by a set of variable $S_i$, where $i$ is the layer index. $S_i$ may take two possible values: +1 if the layer at site ($i$+1) follows the ideal fcc stacking sequence, else a value of $-$1 is assigned. Thus, the energy of a structure with a particular stacking sequence can be expanded as

\begin{eqnarray}\label{eq4}
F&=&J_0-J_1\sum_iS_iS_{i+1}\nonumber\\&-&J_2\sum_iS_iS_{i+2}-J_3\sum_iS_iS_{i+3}+O(J_4),
\end{eqnarray}
where the sums run over the atomic layers. $J_0$ is the energy per unit cell in one layer if the interactions between layers are disregarded. $J_1$, $J_2$, ... are the nearest-neighbor, next nearest-neighbor, \textit{etc.}, interaction parameters between layers. $O(J_4)$ stands for the contribution from the higher order terms. Using this expression, the energies of an intrinsic stacking fault, hcp and double hexagonal close-packed (dhcp) structures relative to the fcc phase can be expressed as

\begin{eqnarray}\label{eq5}
F^{\delta}_{\mathrm{ISF}}-F^0_{\mathrm{fcc}}&=&4J_1+4J_2+4J_3+O(J_4),\nonumber\\
F^{\delta}_{\mathrm{hcp}}-F^0_{\mathrm{fcc}}&=&2J_1+2J_3+O(J_4),\\
F^{\delta}_{\mathrm{dhcp}}-F^0_{\mathrm{fcc}}&=&J_1+2J_2+J_3+O(J_4).\nonumber
\end{eqnarray}
Here the superscript $\delta$ expresses the fact that the corresponding energies are not the equilibrium values. Keeping the interactions up to the third order in Eq. (\ref{eq5}) (\emph{i.e.} neglecting $J_4$ and all higher order terms), for the SFE we get

\begin{equation}\label{eq6}
\gamma^{(3)} = (F^{\delta}_{\rm{hcp}}+2F^{\delta}_{\rm{dhcp}}-3F^0_{\rm{fcc}})/A.
\end{equation}
This third order AIM expression was adopted by Vitos \emph{et al.}~\cite{Vitos2006, Vitos2006acta, Vitos2008, Lu20115728} in the case of paramagnetic steel alloys and it was found to lead to SFE values in good agreement with the experimental results. Truncating Eq. (\ref{eq5}) after the first order (\emph{i.e.} neglecting $J_2$ and all higher order terms), leads to

\begin{equation}\label{eq7}
\gamma^{(1)}= 2(F^{\delta}_{\rm hcp}-F^0_{\rm fcc})/A,
\end{equation}
which is the lowest order approximation for the SFE in terms of interaction parameters.

It is important to realize that when the in-plane and out-of-plane lattice constants are fixed to those of the ideal fcc lattice, $F^0_{\rm fcc}$ in Eqns. (\ref{eq5}-\ref{eq7}) represents the equilibrium fcc total energy, but the other energies do not necessarily correspond to the equilibrium values for the respective structure. That is because Eq. (\ref{eq4}) does not make difference whether an interaction $J_i$ is in fcc, hcp or dhcp matrix. Because of that the AIM cannot account for layer relaxation or magnetic effect which may differ for different packing sequences. This is a shortcoming of the AIM and should be taken into consideration when comparing the AIM predictions with the direct calculations using a supercell technique. In particular, $F^{\delta}_{\rm hcp}$ in Eqns. (\ref{eq5}-\ref{eq7}) represents the total energy of an hcp lattice with $a_{\rm hcp}=\sqrt{2}/2a_0$ and ideal interlayer distance $\lambda_{\rm hcp}=\lambda^0_{\rm fcc}$  (meaning $c_{\rm hcp}/a_{\rm hcp}=\sqrt{8/3}$).

We recall that in Eq. (\ref{eq1}), the SFE is expressed as twice the free energy difference between the equilibrium hcp and fcc phases plus twice the pseudo-interfacial energy $\sigma^{*}$. Thus using Eq. (\ref{eq7}), we arrive at

\begin{equation}\label{eq8}
\gamma = \gamma^{(1)}-2\delta/A+2\sigma^{*} .
\end{equation}
Here $\gamma^{(1)}_0\equiv \gamma^{(1)}-2\delta/A = 2(F^0_{\rm hcp}-F^0_{\rm fcc})/A$ may be taken as the stacking fault energy without fcc/hcp interface energy correction. The correction term in Eq. (\ref{eq8}) is defined as $\delta\equiv F^{\delta}_{\rm hcp}-F^0_{\rm hcp}$. From Eqns. (\ref{eq6}) and (\ref{eq8}) we may infer that when $\delta$ is small, the pseudo-interfacial energy is in fact a correction term from the second and higher order layer interactions to $\gamma^{(1)}$ within the AIM framework. This happens when the equilibrium hcp lattice has similar volume as the fcc lattice and ideal hexagonal axial ratio.

Using the AIM, one may establish a simple relation based on the structural energy differences in Eq. (\ref{eq5}) that can be used to estimate the pseudo-interfacial energy. Making the approximation that $\gamma^{(3)}$ equals the true SFE $\gamma$, from Eqns. (\ref{eq6}) and (\ref{eq7}) we get

\begin{equation}\label{eq9}
\sigma^{*(3)}=(2F^{\delta}_{\rm{dhcp}}-F^{\delta}_{\rm{hcp}}-F^0_{\rm{fcc}})/(2A) +2\delta/A.
\end{equation}
This expression includes interlayer interactions up to the third order ($J_3$). We notice that for most of the metals $F^{\delta}_{\rm{dhcp}}$ is close to the equilibrium dhcp energy $F^0_{\rm{dhcp}}$. The first term in the right hand side of Eq. (\ref{eq9}) gives $\bar\sigma^{* (3)}\equiv\sigma^{* (3)}-2\delta/A$. This quantity may be considered a reasonable approximation to the pseudo-interfacial energy in Eq. (\ref{eq1}) when $\delta$ is negligible.

\subsection{Total energy method}

The total energies were calculated using the exact muffin-tin orbitals (EMTO) method~\cite{Vitos2001, Vitos2007} in combination with the coherent potential approximation (CPA). \cite{Soven1967, Vitos2001b} The EMTO-CPA approach is an appropriate tool for describing systems with chemical disorder. Spin-polarized calculations were performed for Ni. In the self-consistent calculations, the one-electron equations were solved within the scalar-relativistic approximation and soft-core scheme. However, for the three Fe-Cr-Ni alloys, we adopted the frozen-core scheme in order to be able to compare the present results with previous theoretical values \cite{Lu20115728}. The Green function for the valence states was calculated for 16 complex energy points. In the muffin-tin basis set we included $s$, $p$, $d$ and $f$ orbitals. The $k$-point mesh was carefully tested and we used 1000-2000 uniformly distributed $k$-points in the irreducible wedge of the Brillouin zone for fcc, hcp and dhcp structures, and 363 $k$-points for the supercells. With these numerical parameters, the interfacial energy converges within an error of $\sim 1$ mJ m$^{-2}$ (corresponding approximatively to $\sim 0.05$ mRy error in the structural energy differences). For the exchange-correlation functional the generalized gradient approximation of Perdew, Burke, and Ernzerhof was employed. The paramagnetic state of Fe-Cr-Ni alloys were modeled within the disordered local magnetic moment picture. \cite{Gyorffy1985} All calculations were performed at static (0 K) conditions.

\section{Results and discussions}\label{results}

\subsection{Interfacial and stacking fault energies of fcc metals}

Using the present total energy method, first we calculated the equilibrium lattice constants of six fcc metals: Al, Ni, Cu, Ag, Pt, and Au. For each system, the total energies were computed for five volumes around the expected equilibrium value and a Morse type of function \cite{Moruzzi1988} was fitted to extract the equilibrium lattice constant. Results are listed in Table \ref{tab1} together with former theoretical \cite{Shang2010} and experimental values \cite{Kittel1996}. The present predictions are in excellent agreement with previous theoretical values and they are also close to the experimental data.

\begin{table}
\caption{Lattice constant ($a_0$) of fcc structure and the distances between two adjacent fcc(111) ($\lambda^0_{\rm fcc}$) and hcp(0001) ($\lambda_{\rm hcp}$) layers for six selected fcc metals. The hcp lattice was relaxed by constraining the in-plane lattice constant to that of the fcc(111) plane. The present lattice parameters are compared to previous theoretical results and experimental data. Units are \AA.}\label{tab1}
\begin{threeparttable}
\begin{tabular}{ccccccc}
\hline \hline
                        & Al    & Ni     &  Cu    &  Ag    &  Pt  &  Au    \\
 \hline
Present                 & 4.045 & 3.526  & 3.638  & 4.164  & 3.987  & 4.177   \\
Theory\tnote{a}         & 4.046 & 3.518  & 3.631  & 4.148  & 3.980  & 4.164  \\
Expt. \tnote{b}         & 4.05  & 3.52   & 3.61   & 4.09   & 3.92   & 4.08    \\
\hline
$\lambda^0_{\rm fcc} $  & 2.336 & 2.037  & 2.101  & 2.405  & 2.303  & 2.412   \\
$\lambda_{\rm hcp} $    & 2.368 & 2.051  & 2.113  & 2.422  & 2.347  & 2.439   \\
\hline \hline
     \end{tabular}
\begin{tablenotes}
    \item[a] Ref.\cite{Shang2010}
    \item[b] Ref.\cite{Kittel1996}
\end{tablenotes}
\end{threeparttable}
\end{table}

In Table \ref{tab1}, we compare the equilibrium distance between the $(111)$ layers for the fcc lattice ($\lambda_{\rm fcc}^0$) and those between the $(0001)$ layers for the hcp lattice ($\lambda_{\rm hcp}$). These hcp lattices are relaxed by constraining the in-plane lattice parameters to those of the fcc lattices, \emph{i.e.}, $a_{\rm hcp} = \sqrt{2}/2a_0$. For each mono-atomic metal considered here, the equilibrium atomic volume of the fully relaxed hcp lattice is slightly larger than the fcc equilibrium volume and the corresponding equilibrium $c_{\rm hcp}/a_{\rm hcp}$ is also larger than $\sqrt{8/3}$ (ideal value). Thus when stretching the hcp(0001) facet to match the fcc(111) facet, the resulting equilibrium inter-layer distance of the constrained hcp lattice remains slightly larger than the ideal value, \emph{i.e.}, $\lambda_{\rm hcp} >\lambda_{\rm fcc}^0$ (Table \ref{tab1}). In the following, all supercells for interfacial energy calculations are built using the lattice parameters from Table \ref{tab1}, (see Section \ref{superm}).

The calculated interfacial energies for the present elemental fcc metals are listed in Table~\ref{tab2}. We notice that the interfacial energy is very sensitive to the numerical noises and thus special care must be taken when performing the calculations. For all metals considered here, the variations of $\sigma(9,n)$ with increasing number of hcp layers are of the order of our numerical errors. Within these limits, one may conclude that the dependence of the interfacial energy on the number of stacking fault layers is weak. For Cu, Ag and Au, the interfacial energies are very small, the absolute values being close to the present error bars. For Al, Ni and Pt, on the other hand, the absolute values of $\sigma(9,n)$ are surprisingly large. For all metals in Table \ref{tab2}, the converged interfacial energies are negative. Hence, relative to the fcc ground state and the layer-relaxed hcp lattice, the coherent fcc/hcp interfaces are stable.

\begin{table}
\caption{Interfacial energies $\sigma(9,n)$ (in mJm$^{-2}$) for selected fcc metals as a function of the number of hcp layers ($n$) used in the supercell calculations.}\label{tab2}
\begin{tabular}{ccccccc}
\hline \hline
                     &   Al    &   Ni      &   Cu   &   Ag   &   Pt   &    Au   \\
 \hline
   $ \sigma(9,2) $   & -19.367 & -4.413    & -1.275 &  0.009 & -24.052 & -0.847 \\
   $ \sigma(9,4) $   & -21.720 & -6.660    & -0.628 & -0.441 & -24.369 & -1.626 \\
   $ \sigma(9,6) $   & -21.766 & -7.996    & -0.305 & -0.601 & -24.796 & -1.828 \\
   $ \sigma(9,8) $   & -21.766 & -7.996    & -0.305 & -0.601 & -24.796 & -1.828 \\
\hline \hline
\end{tabular}
\end{table}

Considering the lattice mismatch, we find that for all metals in Table \ref{tab2}, the absolute value of the relative difference between the equilibrium hcp lattice parameter in the (0001) plane ($a^0_{\rm hcp}$, not shown) and that of the fcc lattice in the (111) plane ($\sqrt{2}/2a_0$) is below 0.8\%. In particular, for Ni and Ag, the relative difference between the in-plane lattice parameters is around 0.05\%. It means that for these metals the hcp slab in the $(9,n)$ supercells is only weakly deformed relative to the equilibrium hcp lattice. This is reflected in the very small values of $(F^{\rm SC}_{\rm hcp}-F^0_{\rm hcp})$ obtained for Ag and Ni ($\lesssim 50\;\mu$Ry). For Ag, the interfacial energy is close to zero but for Ni we have $\sigma(9,8) \approx -8$ mJm$^{-2}$. In this respect, one may rule out the correlation between the size of the lattice strain at the coherent fcc(111)/hcp(0001) interface and the size of the interfacial energy. On the other hand, the large variation of $\sigma(9,n)$ as we scan the present simple metal (Al), the 3\emph{d} (Ni, Cu), 4\emph{d} (Ag) and 5\emph{d} (Pt, Au) transition metals indicates the importance of the chemistry around the planar fault.

\begin{figure}
	\begin{center}
		\includegraphics[width=8cm]{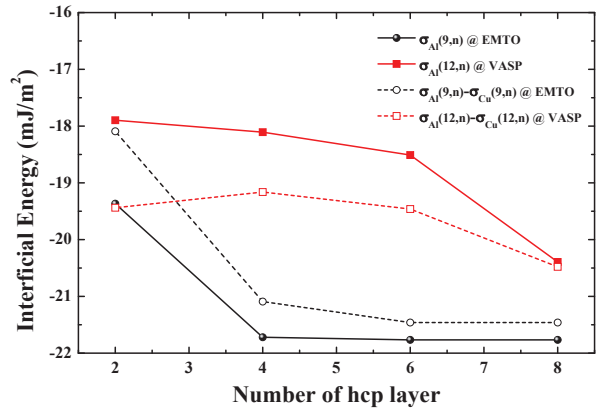}
		\caption{(Color online) Calculated interfacial energies of Al obtained by EMTO and VASP methods. The red (black) solid line with filled square (circle) represents interfacial energy of Al obtained by VASP (EMTO). The dashed red (black) line with open square (circle) stands for interfacial energy difference between Al and Cu, calculated by VASP (EMTO). }
		\label{vasp-int}
	\end{center}
\end{figure}

In order to verify the present predictions for the interfacial energy, we carried out a set of independent supercell calculations for Al and Cu using the Vienna Ab-initio Simulation Package (VASP) \cite{vasp1, vasp2}. In these additional calculations, the generalized gradient approximation (GGA) was considered to describe electronic exchange and correlations \cite{gga}. The interactions between valence electrons and ionic were treated within the projector augmented wave (PAW) basis \cite{paw1, paw2}. The energy cut-off for the plane basis expansion was set to 600 eV for both metals. A 27 $\times$ 27 $\times$ 3 $k$-points mesh with the Monkhorst-Pack scheme \cite{mf} was used to sample the Brillouin zone (BZ) for calculation of $\sigma$(12, n). The structural relaxations, considering atomic position and length of $c$-axis with fixed lateral lattice constant, were performed with the Methfessel-Paxton smearing method \cite{mp} with 0.2 eV smearing width, then the total energy for given structure was obtained by tetrahedron method \cite{th}. Since interfacial energies are very small values, we calculated the total energy with high precision, which was converged to 10$^{-7}$ eV/atoms.

The theoretical equilibrium lattice constants for fcc Al and Cu from VASP calculations are 4.040 and 3.635 {\AA}, which are in good agreement with the present EMTO results. We computed the interfacial energies according to Eq. (\ref{eq2}), with $m=12$ and various $n$. The interfacial energies for Al calculated with VASP and EMTO are presented in Figure ~\ref{vasp-int}. The results obtained with the two methods converge gradually to similar values, with a difference less than 1 mJm$^{-2}$. Since for Cu, the interfacial energy is very small, in Figure ~\ref{vasp-int}, we show the difference $\sigma_{\rm Al}(m,n)-\sigma_{\rm Cu}(m,n)$. Again, the VASP results for Cu are almost identical to the EMTO results. Notice that in VASP calculations, the inter-layer distances are fully relaxed, compared to the local relaxations performed in the EMTO calculations. The agreement between these two sets of data reflects the reliability of the present method.

\begin{table*}
\caption{Theoretical intrinsic stacking fault energies ($\gamma$) and pseudo-interfacial energies ($\sigma^*$) for selected fcc metals (in boldface). The stacking fault energies are the supercell results ($\gamma(9)$) calculated according to Eq. (\ref{eq3}). The pseudo-interfacial energies are calculated from Eq. (\ref{eq1}). The approximate stacking fault energies obtained within the 1st and 3rd order AIM are: $\gamma^{(1)}$ obtained from Eq. (\ref{eq7}), $\gamma^{(3)}$ from Eq. (\ref{eq6}), and $\gamma^{(1)}_0\equiv \gamma^{(1)}-2\delta/A = 2(F^0_{\rm hcp}-F^0_{\rm fcc})/A$. For comparison, former theoretical and experimental stacking fault energies are also shown. Third order approximations $\sigma^{*(3)}$ with and without the $2\delta/A$ term ($\bar\sigma^{* (3)}$) from Eq. (\ref{eq9}) are also shown. The numbers in parentheses for the AIM results are calculated from the corresponding expressions but using the layer-relaxed hcp and dhcp energies with constrained $a_{\rm hcp}$.}\label{tab3}
\begin{threeparttable}
\begin{tabular}{ccccccc}
\hline \hline
   &  Al  &  Ni  &  Cu  &  Ag & Pt &  Au\\
 \hline
  $\gamma$            & \textbf{117.54} & \textbf{153.56}    & \textbf{47.45} & \textbf{17.26}  & \textbf{307.97}   & \textbf{32.69}   \\
  $\gamma^{(1)}$      & 139.28 (136.70) & 160.16 (159.19)    & 50.40 (49.83)  & 17.89 (17.07)   & 359.08 (334.90)   & 34.17 (30.65)    \\
  $\gamma_0^{(1)}$    & 135.53          & 156.92             & 48.62          & 15.80           & 319.83            & 25.22            \\
  $\gamma^{(3)}$      & 123.53 (122.06) & 158.13 (157.41)    & 50.69 (50.48)  & 18.41 (18.06)   & 341.45 (324.68)   & 32.72 (30.50)    \\
  Theory              & 107\tnote{a}    & 153\tnote{a}       & 47\tnote{a}, 56\tnote{b} & 17\tnote{a} , 34\tnote{b}  & 310\tnote{a} , 393\tnote{b}  & 31\tnote{a} , 59\tnote{b}   \\
  Expt.               & 166\tnote{c}   & 125\tnote{d}        & 55\tnote{e}, 41\tnote{f} & 22\tnote{e}, 16\tnote{g}  & 322\tnote{d}     & 50\tnote{e}, 32\tnote{h}       \\
\hline
 $\sigma^{*}$         &  \textbf{-8.99} &  \textbf{-1.68}    & \textbf{-0.58} &  \textbf{0.73}  &  \textbf{-5.93}   &  \textbf{3.73}   \\
\hline
 $ \sigma^{*(3)} $    & -6.00 (-6.74)   &  0.61 (0.24)       &  1.04 (0.93)   &  1.31 (1.13)    & 10.81 (2.42)      &  3.75 (2.64)     \\
 $\bar\sigma^{* (3)}$ &  -7.88 (-7.32)  & -1.01 (-0.89)      &  0.14 (0.32)   &  0.26 (0.49)    & -8.81 (-5.11)     & -0.72 (-0.07)    \\
 \hline \hline
     \end{tabular}
\begin{tablenotes}
    \item[a] Ref.\cite{Li2014}
    \item[b] Ref.\cite{Bcab1993}
    \item[c] Ref.\cite{Murr1973}
    \item[d] Ref.\cite{Hirth1982}
    \item[e] Ref.\cite{Gallager1970}
    \item[f] Ref.\cite{Stobbs1971}
    \item[g] Ref.\cite{Cockayne1971}
    \item[h] Ref.\cite{Jenkins1972}.
\end{tablenotes}
\end{threeparttable}
\end{table*}

Next we turn to the stacking fault energies. The present $(9,2)$ supercells are used in connection with Eq. (\ref{eq3}) to find the stacking fault energy of the present fcc metals. The results are collected in Table \ref{tab2}. Our $\gamma$ values are in good agreement with the previous theoretical results \cite{Li2014,Bcab1993} and also with the limited number of experimental data \cite{Murr1973,Hirth1982,Gallager1970,Stobbs1971,Cockayne1971,Jenkins1972}.

Using the equilibrium hcp ($F^0_{\rm hcp}$) and fcc ($F^0_{\rm fcc}$) total energies, we computed $\gamma^{(1)}_0\equiv 2(F^0_{\rm hcp}-F^0_{\rm fcc})/A$. We recall that $\gamma^{(1)}_0$ is the first term in the right hand side of Eq. (\ref{eq1}) (at 0 K and zero pressure) so that the difference between $\gamma$ and $\gamma^{(1)}_0$ gives twice the pseudo-interfacial energy $\sigma^*$. The results are listed in Table \ref{tab3}. We find that $\sigma^*$ is almost zero for Cu and Ag, and intermediate for Ni and Au. The $\sigma^*$ values for Al and Pt are the largest (in absolute value) among the present metals and both of them are negative. Furthermore, except for Cu and perhaps Ag, the present $\sigma^*$ values differ significantly from the $\sigma(9,n)$ values listed in Table \ref{tab2}. The reason is that the lattice constraint at the fcc(111)/hcp(0001) interface increases the reference hcp total energy substantially relative to the equilibrium hcp energy. The lattice constraint does not influence $\sigma(9,n)$ but pops up directly in $\sigma^*$. That is because in Eq. (\ref{eq1}), $\Delta G^{hcp-fcc}$ refers to the energy difference between two lattices in equilibrium. Thus we suggest that the hcp lattice relaxation effect is primarily responsible for the large difference obtained between $\sigma^*$ and $\sigma(9,n)$. One may expect that ``omitting" the lattice relaxation term would in principle fully remove this disagreement. However, associating the difference between $F^{\rm SC}_{\rm hcp}$ and $F^{0}_{\rm hcp}$ merely to lattice strain given by the coherent interface could be misleading since $F^{\rm SC}_{\rm hcp}$ as extracted from the supercell energies can contain additional terms due to the finite size effects (\emph{e.g.}, interaction between consecutive fcc/hcp interfaces).

For reference, in Table \ref{tab3}, we also list results obtained using the AIM introduced in Section \ref{aimm}. Comparing to the supercell results, the first order AIM ($\gamma^{(1)}$) gives acceptable SFE values for Ni, Cu, Ag and Au (within $\sim$5\% error), but strongly overestimates those for Al and Pt (by more than $\sim$ 15\%). The third order AIM ($\gamma^{(3)}$) somewhat improves the SFE results, especially for Al, but $|\gamma-\gamma^{(3)}|$ for Pt still remains large 33.5 mJm$^{-2}$ (corresponding approximately to $11\%$ error). To reach better agreement for Pt, one may need to keep more terms in Eq. (\ref{eq5}) since previous  $ab$ $initio$ study indicates that the stacking fault in Pt induces longer perturbation in the electronic structures than in other elemental metals.~\cite{Jin2011605}

Sticking to the 3rd order AIM approximation, one can use Eq. (\ref{eq9}) to estimate the pseudo-interfacial energy $\sigma^{*(3)}$. The accuracy of this approximation should be quite high when $\gamma^{(3)}\approx\gamma$. Indeed, the quoted figures in Table \ref{tab3} represent a reasonable approximation for most of the metals considered here: the differences between $\sigma^{*(3)}$ and $\sigma^*$ being close to the errors of the calculations. Exception is Pt, for which $\sigma^{*(3)}$ is surprisingly far from the ``true" $\sigma^*$ value. This is due to the fact that for Pt the 3rd order AIM fails to reproduce the SFE of the supercell model. In Table \ref{tab3}, we also present the AIM estimates for the interfacial energy by excluding the lattice distortion effect (\emph{i.e.}, dropping the $2\delta/A$ term in Eq. (\ref{eq9})). The reason for testing such simplified formula is that within the 3rd order AIM approximation we have $\sigma^{*(3)}-2\delta/A = 2J_2/A$, i.e. the interfacial energy is give by the second order nearest-neighbor layer interaction parameter $J_2$. For most of the metals $2J_2/A$ gives a plausible estimate for the interfacial energy. However, for Au, where the 3rd order AIM is accurate, neglecting $2\delta/A$ in Eq. (\ref{eq9}) worsens the agreement between $\sigma^{*(3)}$ and $\sigma^*$.

%When the metal has small (absolute) value of $\bar\sigma^{* (3)}$, we can classify the metal as "second order" metal because the AIM model up till J$_{2}$ well describes SFE. Conversely, the "high order" metals showing high $\bar\sigma^{* (3)}$ values need higher order in AIM model than others. For instance, the SFE of Ni, Cu, Ag and Au is already close with $\gamma^{(1)}$, but $\gamma^{(3)}$ is require for Al. For Pt case, $\gamma^{(3)}$ is not even enough. Interestingly, the $\gamma^{(3)}$ of Ni and Al are very different although they have high SFE. It is well know that plastic deformation of Ni favors deformation twinning, but full slip dislocation for Al. According to universal scaling law (USL)~\cite{Jin2011605}, which is a model that relates the unstable fault energies and ISF, Pt is totally scattered from the USL line and Al slightly depart from USL~\cite{Li2014,Cai2014}. From this, we can learn that the third order pseudo-interfacial energy is one of the important indicators for understanding plastic deformation behavior. It also reflects that "high order" interaction metals need improved plastic deformation model based on USL with some correction factors.

In application, AIM is often used in connection with hcp and dhcp structural energies obtained for lattices constrained within the $(0001)$ plane but relaxed along the $<111>$ direction. \cite{Lu20115728,Vitos2006acta} We refer to this scheme as the relaxed AIM (r-AIM) approximation. Notice that similar hcp structures are used here to build the inner part of the hcp slab within the $(m,n)$ supercells. To test this scheme for the present systems, in Table \ref{tab3} we list the stacking fault energies and interfacial energies obtained within the r-AIM approximation (numbers in parentheses). Since lattice relaxation lowers the hcp and dhcp energies, the corresponding stacking fault energies also become smaller and eventually closer to the supercell SFE results (except for Au). The interfacial energies obtained from r-AIM $\gamma^{(3)}$ are slightly smaller than the original values. Nevertheless, since relaxation decreases $\delta$, $\sigma^{*(3)}$ and $\sigma^{*(3)}-2\delta/A$ (numbers in parenthesis in the last two rows of Table \ref{tab3}) remain close to each other within the r-AIM scheme.

\subsection{Interfacial and stacking fault energies of Fe$_{\rm 1-x-y}$Cr$_{\rm x}$Ni$_{\rm y}$ alloys}

In this section, we apply the previously developed and tested model to three Fe$_{\rm 1-x-y}$Cr$_{\rm x}$Ni$_{\rm y}$ alloys. We start from a 20\% Cr and 20\%Ni bearing ternary alloy and then first lower the amount of Cr to 13.5\% and then the amount of Ni to 16\%. By that, we can differentiate between the alloying effect of Cr and Ni on the stacking fault energy and interfacial energy of Fe-Cr-Ni alloy.

First we computed the equilibrium lattice parameter and equilibrium interlayer distance for the fcc and constrained hcp structures, respectively. The results are collected in Table \ref{tab4}. We find that within the present approximation, the effects of Cr and Ni on the equilibrium lattice parameters of fcc and hcp structures are negligible (results for hcp not shown). For these alloys, the equilibrium atomic volume of the fully relaxed hcp lattice is slightly smaller than that of fcc and the corresponding equilibrium $c_{\rm hcp}/a_{\rm hcp}$ is also smaller than $\sqrt{8/3}$ (ideal value). Hence, when matching coherently the hcp(0001) facet to fcc(111) facet, the layer-relaxed distorted hcp lattice should have smaller inter-layer separation as compared to the ideal value, \emph{i.e.}, $\lambda_{\rm hcp} <\lambda_{\rm fcc}^0$ (Table \ref{tab4}).

\begin{table}
\caption{Lattice constant ($a_0$) and the distance between two adjacent fcc(111) ($\lambda^0_{\rm fcc}$) and hcp(0001) ($\lambda_{\rm hcp}$) layers for three selected Fe$_{\rm 1-x-y}$Cr$_{\rm x}$Ni$_{\rm y}$ alloys. The hcp lattice was relaxed by constraining the in-plane lattice constant to that of the fcc(111) facet. The present lattice parameters are compared to the theoretical data. Units are \AA.}\label{tab4}
\begin{threeparttable}
\begin{tabular}{ccccccc}
\hline \hline
      & Fe$_{\rm 60}$Cr$_{\rm 20}$Ni$_{\rm 20}$ & Fe$_{\rm 66.5}$Cr$_{\rm 13.5}$Ni$_{\rm 20}$ & Fe$_{\rm 70.5}$Cr$_{\rm 13.5}$Ni$_{\rm 16} $ \\
 \hline
Present   & 3.606  & 3.606  & 3.605   \\
Theory     & 3.599\tnote{a}  & 3.599\tnote{b} & 3.596\tnote{c} \\
\hline
$\lambda^0_{\rm fcc} $  & 2.082  & 2.082  & 2.081   \\
$\lambda_{\rm hcp} $  & 2.056  & 2.056  & 2.056   \\
\hline \hline
     \end{tabular}
\begin{tablenotes}
    \item[a] For Fe$_{\rm 62}$Cr$_{\rm 18}$Ni$_{\rm 20}$ Ref.\cite{Delczeg2012}
    \item[b] For Fe$_{\rm 68}$Cr$_{\rm 12}$Ni$_{\rm 20}$ Ref.\cite{Delczeg2012}
    \item[c] For Fe$_{\rm 72}$Cr$_{\rm 12}$Ni$_{\rm 14}$ Ref.\cite{Delczeg2012}
\end{tablenotes}
\end{threeparttable}
\end{table}

The calculated interfacial energies $\sigma$(9,n) for the three Fe-Cr-Ni alloys are listed in Table~\ref{tab5}. Similarly to the mono-atomic systems, the variations of $\sigma(9,n)$ with increasing number of hcp layers are of the order of our numerical errors. For all three alloys, the converged interfacial energies are small and negative, with absolute values larger than the present error bars. Considering $\sigma(9,n)$ with $n\geq 4$, one may conclude that both Cr and Ni addition to Fe-Cr-Ni alloys increases the interfacial energy. However, these alloying induced changes in $\sigma(9,n)$ are very small and thus the trends should be considered with precaution.

\begin{table}
\caption{Interfacial energies $\sigma(9,n)$ (in mJm$^{-2}$) for three selected Fe$_{\rm 1-x-y}$Cr$_{\rm x}$Ni$_{\rm y}$ alloys as a function of the number of hcp layers ($n$) used in the supercell calculations.}\label{tab5}
\begin{tabular}{cccc}
\hline \hline
 & Fe$_{\rm 60}$Cr$_{\rm 20}$Ni$_{\rm 20}$ & Fe$_{\rm 66.5}$Cr$_{\rm 13.5}$Ni$_{\rm 20}$ & Fe$_{\rm 70.5}$Cr$_{\rm 13.5}$Ni$_{\rm 16} $ \\
 \hline
    $\sigma(9,2)$      &  0.985 &  0.465 &  0.709  \\
    $\sigma(9,4)$      & -0.873 & -1.644 & -1.693  \\
    $\sigma(9,6)$      & -2.132 & -2.535 & -2.972  \\
    $\sigma(9,8)$      & -2.132 & -2.535 & -2.972  \\
 \hline \hline
 \end{tabular}
\end{table}

The present $(9,2)$ supercells are used in connection with Eq. (\ref{eq3}) to find the stacking fault energies of Fe-Cr-Ni alloys. The results are listed in Table \ref{tab6}. Our $\gamma$ values are in good agreement with the previous theoretical results \cite{Lu20115728,Vitos2006} and also with the limited number of experimental data. \cite{Kaneko,Fawley,Brofman} Chromium is found to decrease the SFE and Ni slightly increase it, which is in line with the previous theoretical predictions.\cite{Vitos2006,Vitos2006acta,Lu20115728}

\begin{table}
\caption{Theoretical intrinsic stacking fault energies ($\gamma$) and pseudo-interfacial energies ($\sigma^*$) for selected Fe$_{\rm 1-x-y}$Cr$_{\rm x}$Ni$_{\rm y}$ alloys. For notations see caption for Table \ref{tab3}.}\label{tab6}
\begin{threeparttable}
\begin{tabular}{ccccccc}
\hline \hline
   & Fe$_{\rm 60}$Cr$_{\rm 20}$Ni$_{\rm 20}$ & Fe$_{\rm 66.5}$Cr$_{\rm 13.5}$Ni$_{\rm 20}$ & Fe$_{\rm 70.5}$Cr$_{\rm 13.5}$Ni$_{\rm 16} $ \\
 \hline
  $\gamma$              & \textbf{30.55} & \textbf{39.38} & \textbf{38.22}  \\
  $\gamma^{(1)}$        & 21.65 (15.19) & 30.31 (25.32) & 28.94 (23.09) \\
  $\gamma_0^{(1)}$      & 14.95 & 22.84 & 20.61  \\
  $\gamma^{(3)}$        & 28.94 (25.96) & 37.46 (34.96) & 35.97 (33.01) \\
  Theory   &  35\tnote{d}& 50\tnote{d} &45\tnote{d}  \\
  Expt.     &   32\tnote{a}, 38\tnote{b}, 41\tnote{c}& 46\tnote{c}  &38\tnote{c}  \\
\hline
 $\sigma^{*}$           & \textbf{7.80} &  \textbf{8.27} &  \textbf{8.80}  \\
\hline
 $ \sigma^{*(3)}       $&  6.99 (5.51) &  7.31 (6.06) &  7.68 (6.20)  \\
 $\bar\sigma^{* (3)}$&  3.65 (5.14) &  3.57 (4.82) & 3.52 (4.96)  \\
 \hline \hline
     \end{tabular}
\begin{tablenotes}
\item[a] Ref.\cite{Kaneko} \item[b] Ref.\cite{Fawley} \item[c] According to SFE=16.7+2.1Ni-0.9Cr+26C (mJm$^{-2}$) from Ref.\cite{Brofman}. \item[d] Refs.\cite{Vitos2006,Vitos2006acta}.
\end{tablenotes}
\end{threeparttable}
\end{table}

Table \ref{tab6}, also shows the stacking fault energies obtained within the AIM and r-AIM (numbers in parentheses). It is interesting to notice that for the present mono-atomic systems (Table \ref{tab3}) both the 1st and 3rd order AIM SFE values ($\gamma^{(1)}$ and $\gamma^{(3)}$) are larger than $\gamma$. For Al, Ni, Cu and Pt, the relaxed 1st order approximation ($\gamma_0^{(1)}$) remains also above $\gamma$. The situation is very different for the steel alloys in Table \ref{tab6}, for which we have $\gamma_0^{(1)}<\gamma^{(1)}<\gamma^{(3)}<\gamma$. This trend is reflected in the large positive pseudo-interfacial energies $\sigma^*$ obtained for all three alloys. Both Ni and Cr additions seem to decrease the pseudo-interfacial energy, although the variations are very small. Our predict pseudo-interfacial energies are slightly smaller than the values proposed by Olson and Cohen ($\sigma^{*}$=10-15 mJ m$^{-2}$). \cite{Olson1976}

For the present paramagnetic steel alloys, the 3rd order AIM performs reasonably well. Namely, the absolute difference between $\gamma^ {(3)}$ and $\gamma$ is less than $\sim$6\%, and most importantly, the alloying effects on the SFE predicted by the two methods are the same. In consequence, the interfacial energies $\sigma^{*(3)}$ computed from the 3rd order AIM expression Eq. (\ref{eq9}) are also close to $\sigma^*$. It is important to note that the trend of $\sigma^*$ is also well reproduced by the 3rd order AIM.  However, when we omit the lattice distortion effect ($2\delta/A$), the performance of AIM for predicting the pseudo-interfacial energy is to large extend destroyed.

Unlike for the mono-atomic systems in Table \ref{tab3}, in the case of steel alloys adopting the r-AIM scheme does not improve the agreement between the supercell and AIM results. The resulted composition dependence of the SFE is the same, but the corresponding SFE values are further underestimated compared to the supercell results.

\section{Conclusions}

First-principles calculations have been performed to determine the stacking fault energy and the fcc(111)/hcp(0001) interfacial energy for six elemental metals and three paramagnetic Fe-Cr-Ni alloys, which form the basis building block of austenitic stainless steels. For all systems, the predicted SFE results are in good agreement with the previous theoretical values and the available experimental data, confirming the accuracy of the present approach. The calculated interfacial energy, however, depends strongly on the reference structures. We define the interfacial energy that enters the thermodynamical SFE calculations as the pseudo-interfacial energy ($\sigma^ {*}$) with the equilibrium hcp and fcc structures as reference states. $\sigma^ {*}$ is calculated according to the thermodynamical expression for SFE (Eq. (\ref{eq1})) using the calculated SFE from our supercell model. The results show that the coherent interfacial energy differs significantly from the pseudo-interfacial energy.

For all elemental metals and Fe-Cr-Ni alloys considered here, the coherent interfacial energy ($\sigma(9,n)$) defined as the excess energy at the coherent interface between semi-infinite fcc and hcp slabs is small or negative. On the other hand, the pseudo-interfacial energy shows a large variation as we go from simple metal Al, to transition metals (Ni, Cu, Ag, Pt and Au) and transition metal alloys (Fe-Cr-Ni), which indicates  that the $\sigma^ {*}$ strongly depends on chemistry. Our predicted pseudo-interfacial energies for the Fe-Cr-Ni alloys agree well with commonly quoted values in the literature.

The accuracy of the often employed Axial Interaction Model has also been scrutinized. Whenever the 3rd order AIM performs well for SFE it also predicts the pseudo-interfacial energy $\sigma^*$ with sufficient high accuracy. This is the case for Fe-Cr-Ni alloys, Cu, Ag and Au, and to lesser extend for Al and Ni. However, the 3rd order AIM fails for Pt, which might be cured by including further axial interactions to capture the large negative $\sigma^*$ value. No significant improvement is found when comparing the relaxed-AIM to the original AIM approximation.

\section*{Acknowledgments}

The Swedish Research Council, the Swedish Foundation for Strategic Research, the Carl Tryggers Foundation, the Chinese Scholarship Council, the National Magnetic Confinement Fusion Energy Research Project of China (2015GB118001) and the Hungarian Scientific Research Fund (OTKA 84078 and 109570) are acknowledged for financial support. Song Lu thanks the Magnus Ehrnrooth foundation for providing a Postdoc. Grant. We acknowledge the Swedish National Supercomputer Centre in Link\"oping for computer resources.

\end{document}